\documentclass[aps,prb,superscriptaddress,reprint]{revtex4-1}

\PassOptionsToPackage{elide}{natbib}
\usepackage{graphicx} 
\usepackage{bm}
\usepackage{amssymb,amsmath}

\newcommand{\tr}{\operatorname{tr}}

\begin{document} 

\title{Stability of fractional vortex states in a two-band mesoscopic superconductor}

\author{Juan C. Pi\~na}
\affiliation{Departamento de F\'isica, Universidade Federal de Pernambuco, Cidade Universit\'aria, 50670-901, Recife-PE, Brazil}
\author{Cl\'ecio C. de Souza Silva} 
\affiliation{Departamento de F\'isica, Universidade Federal de Pernambuco, Cidade Universit\'aria, 50670-901, Recife-PE, Brazil}
\author{Milorad V. Milo\v{s}evi\'{c}}
\affiliation{Departamento de F\'isica, Universidade Federal do Cear\'a, 60455-900 Fortaleza, Cear\'a,
Brazil}
\affiliation{Departement Fysica, Universiteit Antwerpen, Groenenborgerlaan 171, B-2020 Antwerpen, Belgium}

\date{\today}

\begin{abstract} 
We investigate the stability of noncomposite fractional vortex states in a mesoscopic two-band superconductor within the two-component Ginzburg-Landau model. Our analysis explicitly takes into account the relationship between the model parameters and microscopic material parameters, such as partial density of states, fermi velocities and elements of the electron-phonon coupling matrix. We have found that states with different phase winding number in each band ($L_1\neq L_2$) and fractional flux can exist in many different configurations, including rather unconventional ones where the dominating band carries larger winding number and states where $|L_1-L_2|>1$. We present a detailed analysis of the stability of the observed vortex structures with respect to changing the microscopic parameters, showing that, in the weak coupling case, fractional vortex states can be assessed in essentially the whole range of temperatures and applied magnetic fields in which both bands are active. Finally, we propose an efficient way of increasing the range of parameters for which these fractional vortex states can be stabilized. In particular, our proposal allows for observation of fractional vortex structures in materials with stronger coupling, where those states are forbidden at a homogeneous field. This is accomplished with the help of the stray fields of a suitably prepared magnetic dot placed nearby the superconducting disk.
\end{abstract}

\maketitle

\section{Introduction}

Many properties of the electronic condensate in a macroscopic quantum system can be unveiled by rotating it and studying the structure of the induced quantized vortices.\cite{Leggett2008} In particular, the amount of magnetic flux carried by Cooper-pair vortices in bulk conventional superconductors determines whether the material is of type I or type II.\cite{Abrikosov57} In the first case, vortices tend to merge into large flux domains carrying many flux quanta. Conversely, for the type-II material, vortices are singly quantized and repel each other, tending to arrange themselves into a triangular lattice.

Some superconductors, however, are provided with two or more electronic condensates arising from Cooper pairing in different bands of the material. These multiband systems have attracted much interest in the last decade because: (i) most of the recently discovered superconducting materials, such as MgB$_2$ and compounds of the iron-pnictide family, are recognized as multiband superconductors; and (ii) they exhibit a variety of new and interesting phenomena with no counterpart in conventional single-component superconductors. Perhaps the most intriguing of these phenomena are related to the exotic vortex structures that can emerge in a multicomponent superconductor. 

In the bulk, a multicomponent vortex can exist in equilibrium only in a so-called composite state where vortices in the different condensates share the same core.\cite{Babaev2002} However,  because of the different length scales $\xi_i$ at which the Cooper-pair density varies in each component,\cite{Komendova2011,Silaev2012} the interaction between such composite vortices can be non-monotonic: short-range repulsive with an attractive tail.\cite{Silaev2011} This ultimately leads to vortex clustering at low magnetic fields and, thereby, to the formation of a semi-Meissner state, that is, a mixture of flux-free and vortex-cluster regions coexisting in equilibrium. This phase was predicted by Babaev and Speight\cite{Babaev2005} and experimentally observed by Moshchalkov {\it et al.}\cite{Moshchalkov2009} in an ultraclean MgB$_2$ single crystal. In this case, the first critical field corresponds to the thermodynamic stabilization of a vortex cluster in the sample, rather than a single vortex. In this regime, besides clusters, other
exotic configurations are likely, such as vortex rings, giant-multi
vortex groupings, and other unusual
patterns.\cite{Moshchalkov2009,Dao2011}

Actually, the competing interactions in a multiband material can
lead to even more unconventional vortex topologies. For instance, ``unmatched'' composite vortices, that is, with both condensates having different phase winding numbers, can exist out of equilibrium.\cite{Babaev2009,ChoZhang2008} Such vortices enclose arbitrary magnetic flux, in contrast to quantized Abrikosov vortices in conventional superconductors. Another example are noncomposite vortices, in which the phase singularities in each condensate are displaced from one another. These vortices also carry arbitrary flux, however, since their energy increases prohibitively with sample size, they can only be observed in small samples. These exotic vortex structures could be used as a hallmark of multiband superconductivity. Therefore,  an important question is how to stabilize and detect such fractional vortices. 

Recently, Chibotaru and co-workers\cite{Chibotaru2007,Chibotaru2010} have proposed that fractional vortices can be realized and even thermodynamically stabilized in a two-band mesocopic disk. Subsequently, Geurts {\it et al.}\cite{Geurts2010} extended the analysis by including magnetic coupling, besides the interband Josephson coupling used in previous works. All these studies were performed within the two-component Ginzburg-Landau (TCGL) formalism, taking phenomenological constants as parameters of the simulations. However as recently observed in Ref.~\onlinecite{Chaves2011}, the phenomenological constants of the TCGL model cannot be chosen freely because they are coupled by microscopic material parameters, such as electron-phonon coupling constants, Fermi velocities and partial density of states. A more systematic study of the stability of composite and non-composite vortex states in mesoscopic two-component superconductors, within the correct microscopic framework, is still pending.

In this work, we investigate the stability regions of fractional flux vortices in a two-band mesoscopic superconductor in a parameter space defined by microscopic material parameters. In addition, we demonstrate how the stability of these states can be considerably enhanced by the presence of a closeby magnetic dot with a suitable magnetization.

\section{Theoretical formalism and simulation details}
\label{sec.model}

We consider a mesoscopic two-band superconducting disk of thickness $d$ much smaller than the penetration depth $\lambda$ and the two characteristic lengths of density variations in both condensates, in such a way that the system is effectively two-dimensional. Our calculations rely on the minimal two-component Ginzburg-Landau (TCGL) model, where the only coupling appearing explicitly in the free-energy functional is Josephson-like.\cite{Zhitomirsky2004,Chibotaru2007,Chibotaru2010,Geurts2010,Chaves2011} Within this framework, the total free energy can be written as the sum of the free energies of the otherwise isolated bands and the Josephson coupling free energy, that is,
\begin{eqnarray}
 F &=& \int \! d^3r \Bigg\{\sum_{j}\bigg[\frac{1}{2m_j}\left|\left(-i\hbar\nabla-\frac{2e}{c}\vec{A}\right)\Psi_j\right|^2 \nonumber \\
 & & + \alpha_j|\Psi_j|^2  + \frac{1}{2}\beta_j|\Psi_j|^4 \bigg] - \Gamma\left(\Psi_1\Psi_2^* + \Psi_1^*\Psi_2\right)\!\Bigg\}
 \label{eq.freeenergy}
\end{eqnarray}
$j=1,2$ is the band label and $m_j$, $\alpha_j$, $\beta_j$, and $\Gamma$ are parameters derivable from the microscopic theory. The last term accounts for magnetic coupling between the bands,
where $\vec{H}_0$ is the applied magnetic field, and
$\vec{h}=rot\vec{A}$ is the local (total) field. However, because of the small thickness and lateral size of the system under investigation, screening effects are negligible and thereby so is the magnetic coupling between the bands.

Within the two-band Eilenberger formalism,\cite{Kogan2011,Silaev2011,Silaev2012} the relevant microscopic parameters are the Fermi velocities $v_j$, the partial densities of states $n_jN(0)$, and the elements of the electron-phonon coupling matrix
\begin{equation}
\Lambda = \left( \begin{array}{cc}
n_1\lambda_{11} & n_2\lambda_{12} \\
n_1\lambda_{12} & n_2\lambda_{22}  
\end{array} \right). \nonumber
\end{equation}
The critical temperature is given by the relation $1.76T_c=2\hbar\omega_De^{-S}$, where $\omega_D$ is the Debye frequency and
\begin{equation}
 S=\frac{\tr\Lambda \pm \sqrt{(\tr\Lambda)^2 - 4\det\Lambda}}{2\det\Lambda}. \nonumber
\label{eq.S}
\end{equation}
are the roots of the linear system of self-consistecy equations for the gaps. The correct value of $T_c$ is given by the smallest root. It is also convenient to define the following auxiliary parameters: $\tau=-\ln(T/T_c)\simeq 1-T/T_c$, $W=2\sqrt{2}\pi T_c/\sqrt{7\zeta(3)}$, $\eta=\det \Lambda/n_1n_2$, and the positive constants
\begin{equation}
 S_1=\lambda_{22}-n_1\eta S \qquad \text{and} \qquad S_2=\lambda_{11}-n_2\eta S. \nonumber
\label{eq.Sj}
\end{equation}

In terms of the microscopic parameters above, the TCGL coefficients can be expressed as: $\alpha_j=-N(0)n_j\left(\tau-{S_j}/{n_j\eta}\right)$, $\beta_j={N(0)n_j}/{W^2}$, $m_j={3W^2}/(N(0)n_jv_j^2)$, and $\Gamma={N(0)\lambda_{12}}/{\eta}$. As usual, one can define the healing lengths $\xi_j=\hbar v_j/\sqrt{6W}$, which are related to the characteristic lengths of density variations of order parameters 1 and 2 (see for instance Ref.~\onlinecite{Komendova2011}). However, the real healing lengths of the condensates have to be calculated, and are strongly influenced by the critical temperatures of the otherwise uncoupled bands\cite{Komendova2012} (that is, the temperature at which the corresponding $\alpha_j$ changes sign),
\begin{equation}
 T_{cj} = T_c\exp\left(-\frac{S_j}{n_j\eta}\right) 
 \simeq T_c\left(1 - \frac{S_j}{n_j\eta} \right).
 \label{eq.Tcj}
\end{equation}
Notice that both $T_{cj}$ are always smaller than $T_c$. Therefore, for $T_{c1},T_{c2}<T<T_c$, both $\alpha_j$ are positive and superconductivity survives in the system only due to coupling between the bands, whereas for $T<T_{c1},T_{c2}$, both bands are active ($\alpha_j<0$). For temperatures such that $T_{c\textrm{P}}<T<T_{c\textrm{A}}$, one of the bands (band A) is active while the other (band P) is passive, that is, it remains superconducting only because of Cooper pairs coming from band A.

Finally, minimization of the TCGL free energy~\ref{eq.freeenergy} leads to the following dimensionless GL equations:
\begin{subequations}
\begin{align}\label{eq.TCGL}
 (-i\nabla-\vec{A})^{2}\psi_1 - (\tau_1 -  |\psi_1|^2)\psi_1- 
 \frac{\lambda_{12}}{n_1\eta}\psi_2=0, \\
 \frac{v_2^2}{v_1^2}(-i\nabla-\vec{A})^{2}
 \psi_2-(\tau_2-|\psi_2|^2)\psi_2-
 \frac{\lambda_{12}}{n_2\eta}\psi_1=0,
\end{align}
\end{subequations}
where $\tau_j \equiv -\ln(T/T_{cj}) = \tau - S_j/(n_j\eta)$. Here, we adopted the following temperature independent units: $T_c$ for temperatures, $W$ for both order parameters, $\xi_1$ for distances, and $A_0=\hbar c/2e\xi_1$ for the vector potential. Hereafter, we choose the disk radius $R=10\xi_1\ll\lambda_{\textrm{eff}}$, where $\lambda_{\textrm{eff}}=\lambda^2/d$ is the effective penetration depth, so that screening and demagnetizing effects can be neglected. Accordingly, the vector potential in (\ref{eq.TCGL}) can be well approximated by that corresponding to the external magnetic field.

The minimal TCGL model with the $\tau\simeq 1-T/T_c$ approximation is in principle strictly valid only in the immediate vicinity of $T_c$, in which case the model is reduced to the conventional GL theory featuring a single order parameter with a single length scale.\cite{Kogan2011} However, for lower temperatures the correct microscopic scenario points to two condensates with two {\it a priori} different healing lengths. Inspired by the success of the GL formalism in producing correct results well out of its strict validity range in several other instances, one may ask whether the applicability range of the TCGL model can be pushed to lower temperatures. Indeed, in Ref.~\onlinecite{Shanenko2011} it was shown that an extension of the GL theory to include complete $\tau^{3/2}$ terms to the expansion of the order parameters pushes the validity of the theory to temperatures as low as $0.62T_c$, depending on the values of the microscopic parameters. In addition, this model correctly predicts two different density length scales, which become equal only for $T\rightarrow T_c$, thus in reconciliation with previous results from the microscopic theory. Unfortunately, the extended GL model for the non-zero field case is to date available only for the single-band superconductors.\cite{Vagov2012}

Recently, evidences that the standard TCGL model provides a correct description of two-band superconductors has been given in Ref.~\onlinecite{Silaev2012}. The authors set out a systematic comparison with the microscopic two-band Eilenberger theory suggesting that the minimal TCGL model can offer a quite accurate \emph{quantitative} description of features specific of two-band superconductivity, including two-length-scales vortex solutions, if one retains the full logarithmic temperature dependence of $\tau$ for temperatures smaller but not too close to $T_c$. Good agreement was achieved for temperatures as low as $T=0.85$. Therefore, to ensure the validity of the TCGL model used in the present work, we fix the temperature of the sample at $T=0.90$ and choose microscopic parameters such that both $T_{c1}$ and $T_{c2}$ are also close to $T_c$ (typically, higher than 0.9).  Such critical temperatures are also chosen in order to minimize the effects of hidden criticality in the case of weak coupling between the bands (see Ref. \cite{Komendova2012} for details).

The calculations are performed as follows. Eqs.~\ref{eq.TCGL} are numerically integrated by a relaxation method where a diffusion-like time-dependence of the fields is assumed. This procedure is similar to solving the time-dependent Ginzburg-Landau equations. However, once we are interested only in the stationary states, we chose the same diffusion constant, $D=1$, for both order parameters. The equations are then discretized in space following the link-variable method and the resulting gauge-invariant, finite-difference equations are time-integrated via a semi-implicit scheme.\cite{Winiecki2002} As compared to the conventional Euler method, the semi-implicit integration provides better stability and higher accuracy, which are particularly crucial for the convergence of fractional vortex solutions. The results are presented as functions of the external magnetic flux threading the disk area, $\Phi=H/\pi R^2$, in units of the magnetic flux quantum, $\Phi_0=hc/2e$. The configurations corresponding to bands 1 and 2 having a phase winding number $L_1$ and $L_2$, respectively, are labeled as $(L_1,L_2)$.

\section{Influence of microscopic parameters}\label{sec.micro}

In most previous studies on fractional vortex states, the main focus of analysis was on specific material parameters. However, to date there is no consensus about the correct coupling constants of, for instance, MgB$_2$, by far the most studied two-band material (see for instance Ref.~\onlinecite{Brandt2011} and references therein). Therefore, here we adopt a different strategy and study the possible configurations of Cooper pair densities in different regions of the parameter space defined by the microscopic quantities $\lambda_{12}$, $n_1$ ($=1-n_2$) and $v_2$. Notice that in our calculations $v_1$ is a constant bound to the unit length. For a treatable analysis, we fix only the intraband coupling constants, $\lambda_{11} = 2.415$ and $\lambda_{22}=1.211$. The only bias in such a choice is that these values lead to $T_{c1}$ and $T_{c2}$ close to $T_c$, which is important for the applicability of the Ginzburg-Landau framework. However, we believe that other values would impart no qualitative change to the main conclusions of our investigation. 

\subsection{Effect of the partial densities of states}\label{ssec.n1}

Here, we fix the interband coupling at $\lambda_{12}=0.001$ and the Fermi velocity ratio at $v_1/v_2=1.225$ and change only the partial densities of states, $n_1$ and $n_2$. The main effect of changing these quantities is the relative variation of the critical temperatures of band activity, $T_{c1}$ and $T_{c2}$. Although our choice of intraband coupling constants (with $\lambda_{11}\simeq2\lambda_{22}$) benefits band 1 as, in principle, the strongest band, one can always have $T_{c2}>T_{c1}$ if $n_1$ and $n_2$ are properly chosen. In fact, it is the product $n_j\lambda_{jj}$ which determines which of the bands will dominate. It results from Eq.~\ref{eq.Tcj} that both bands has the same $T_{cj}$ when $n_1\lambda_{11} = n_2\lambda_{22}$. For our choice of the coupling matrix, this reads $n_1 = 1-n_2 = 0.334$. Band 1 is the dominating band above this line only, whereas band 2 dominates below it. By moving away from the $T_{c1}=T_{c2}$ line, in any direction, one of the bands will become passive. This happens with band 1 for $n_1=1-n_2<0.317$ and with band 2 for $n_1=1-n_2>0.354$. 

\begin{figure}[tb]
\centering
  \includegraphics[width=\columnwidth]{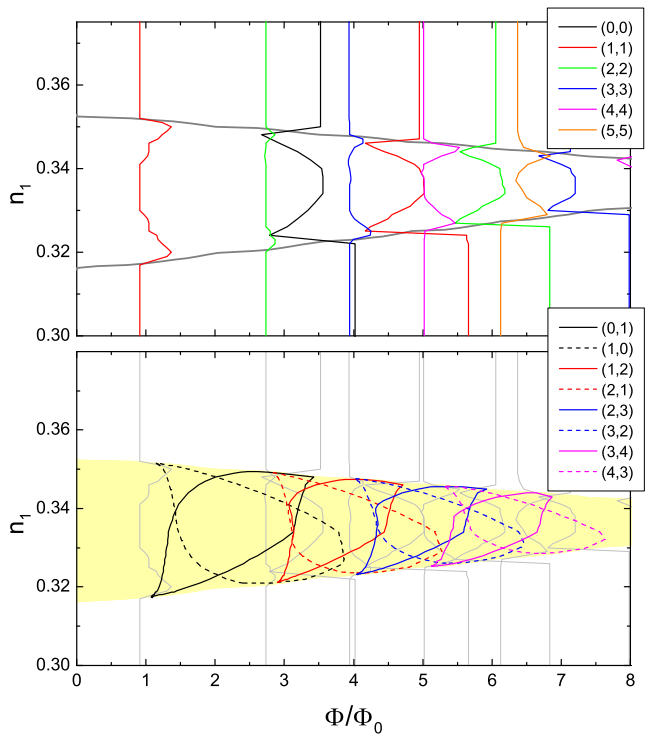}
  \caption{(Color online.) Stability regions of integer and fractional $(L_1,L_2)$ configurations in the plane defined by the partial density of states $n_1$ and reduced external magnetic field $\Phi/\Phi_0$. Integer $(L,L)$ states are bounded by two full lines, the left (right) one corresponding to the smallest (highest) field value below (above) which the state becomes unstable. The stability regions of fractional states correspond to the closed regions delimited by full lines for $L_1<L_2$ and dashed lines for $L_1>L_2$. For comparison, we show the stability limits of integer states as light gray lines. The heavy gray lines in the upper panel delimits the region (shaded area in the lower panel) where both bands are active.}
  \label{fig.n1xPhi} 
\end{figure}

Fig.~\ref{fig.n1xPhi} 
presents the stability regions of $(L_1,L_2)$ states in the $n_1$-$\Phi$ plane. The curves plotted in these diagrams correspond to the limit of stability of each particular $(L_1,L_2)$ state, as explained in the caption. For a given $n_1$, each stationary state was obtained after initializing the system with a particular $(L_1,L_2)$ configuration at a given value of $\Phi$ and checking whether it converges to a stationary solution. Once a field value was found in which that configuration is stable, we swept the field up and down to check the whole range of stability. As expected, the appearance of fractional states, i.e., states with $L_1\neq L_2$, is only possible in the range of $n_1$ and $n_2$ values for which both bands are active. Outside this region, the active band forces the passive band to follow its configuration via Josephson coupling. Surprisingly, both combinations $L_1<L_2$ and $L_1>L_2$ appear as stable $(L_1,L_2)$ states in most part of the two-active-band region, no matter which of the two bands has a higher $T_{ci}$. This is in contrast with previous studies, where only states with $L_1\leq L_2$ (band 1 being the strongest band) were reported. 

\begin{figure}[bt]
\centering
  \includegraphics[width=\columnwidth]{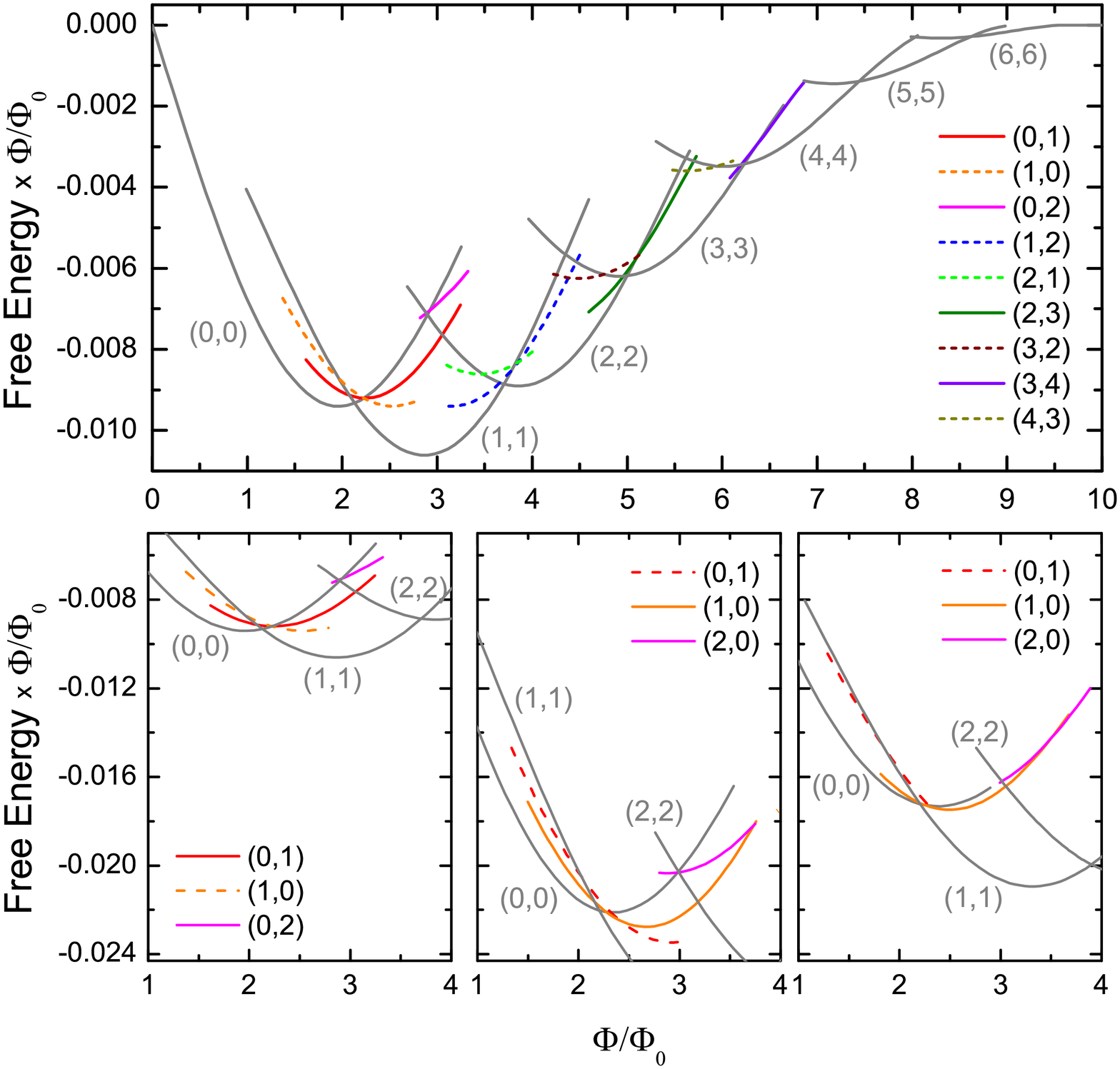}
  \caption{(Color online.) Free energy as a function of external
magnetic field (flux). Full lines correspond to states stabilized by
sweeping the field up and down starting from the Meissner state at
$H=0$ as initial condition (see text). Dashed lines correspond to
the states obtained only by imposing a suitable initial condition,
as explained in the text. Top: all states found for the case
$n_1=0.343$. Bottom: selected fractional states [(0,1), (1,0), (0,2) and (2,0)] for three different values of the partial density of states: $n_1=0.343$ (left), 0.334 (middle), and 0.325 (right).}
  \label{fig.FxPhi} 
\end{figure}

A detailed analysis of how these states change with field and how they are energetically compared to one another is presented in Fig.~\ref{fig.FxPhi}. 
The results shown in the upper panel of this figure were calculated for $n_1=0.343$, for which $T_{c1}=1.0$ and $T_{c2}=0.95$. Therefore, the ``inverted'' states are those with $L_1>L_2$. The variety of possible configurations in a two-band material is immediately revealed. For instance, at a field $\Phi=3.2\Phi_0$ we found as much as \emph{seven} different stable configurations. It
is also clear in this figure that all fractional states are
metastable, i.e. have higher energy than the ground state. Note that
fractional states can be also found in the ground state, but for
very weak interband coupling, as discussed later in
Sec.~\ref{ssec.coupling}. In the present case, the states with
$L_1<L_2$ have, in general, lower energy, sometimes very close to
the ground states. An exception is the (0,2) state, with a remarkably high energy. The evolution of relative energies and stability range of a selection of fractional states as one decreases $n_1$ is illustrated in the lower panel of Fig.~\ref{fig.FxPhi}. As expected, for $n_1=0.325$ ($T_{c1}=0.95$ and $T_{c2}=1.0$) the roles of states with $L_1>L_2$ and those with $L_1<L_2$ are interchanged, with the former having in general lower energy. 

Another important feature distinguishing ``regular'' and ``inverted'' $(L_1,L_2)$ solutions concerns the way these states are found. The curves in Fig.~\ref{fig.FxPhi} represented by full lines were obtained after initializing the system in either the Meissner state, at $H=0$, or normal state, at $H$ just above the sample's upper critical field, and then isothermally sweeping the field up and down. Such a procedure accounts for a realistic magnetic history and  allows one to access both metastable and ground states. Experimentally, it is similar to performing minor isothermal magnetization-loops measuremets on mesoscopic samples\cite{Geim97,Geim98}. In this case, finding a new vortex state involves entry or exit of one or more vortices through the surface barrier once a saddle point is reached.\cite{Schw1999,Palacios2000,Baelus2001} Therefore, we anticipate that the fractional states represented by full lines in Fig.~\ref{fig.FxPhi}, though metastable, should be easily accessed experimentally. On the other hand, those states represented by dashed lines could not be obtained this way. Instead, we used a particular vortex configuration as the initial state, in anticipation of a similar stable outcome of the calculation. We are not aware of an experimental counterpart of this process. One could then conclude that, although inverted fractional states do appear as stable solutions of the Ginzburg-Landau equations, they are unlikely to be observe experimentally.

\subsection{Effect of the Fermi velocities}\label{ssec.fermi}

\begin{figure}[t]
\centering
  \includegraphics[width=0.9\columnwidth]{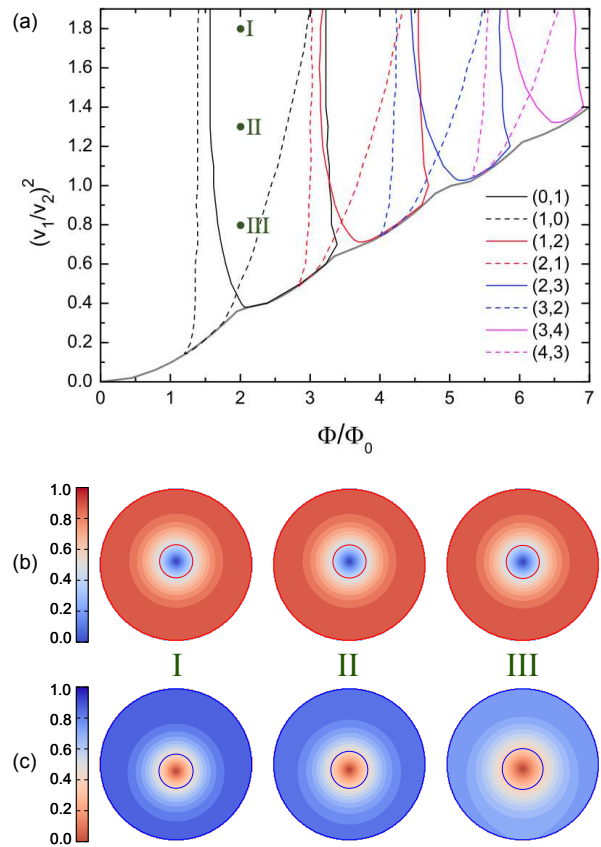}
  \caption{(Color online) (a) Stability regions of fractional $(L_1,L_2)$ states in the plane defined by the ratio between the Fermi velocities $(v_1/v_2)$ and the applied magnetic flux $\Phi$. The gray line corresponds to superconducting-normal phase boundary of band 2 in the absence of interband coupling. (b) and (c): Plots of the absolute value of the order parameter of band 1 in the (1,0) state (b) and band 2 in the (0,1) state (c) for the point I, II, and III indicated in (a). The order parameters are normalized by the maximum zero-coupling value of each band. The circles indicate the size of the vortex as defined in the text.}
  \label{fig.vxPhi} 
\end{figure}

In this subsection, we retain the coupling constants defined in the previously, fix $n_1=1-n_2=0.343$, and analyze the effect of Fermi velocities on the vortex states in a two-band mesoscopic disk. As
shown in the theoretical formalism, the Fermi velocity of a band is
directly proportional to the healing length of the corresponding
condensate at zero temperature. In other words, $v_j$ is closely connected to the maximum critical field band $j$ can sustain in the zero coupling limit and at $T=0$. Bearing in mind that $\xi_1$ is fixed as our unit length (and so $v_1$ remains fixed as well), by changing the ratio $v_1/v_2$ as a parameter we are actually changing $\xi_2$. Therefore the critical field above which band 2 becomes passive is expected to increase with $(v_1/v_2)^2$. Fig.~\ref{fig.vxPhi} 
presents the evolution of the stability regions of fractional states as a function of the squared Fermi velocity ratio $(v_1/v_2)^2$. In addition, we plot the phase boundary separating the active and passive states of band 2. Above this line, both bands are active. The critical field for band 1 activity (not shown) corresponds to the constant value $9.5\Phi_0$. 

The role played by the Fermi velocities and the healing lengths in determining the shape of the stability regions in Fig.~\ref{fig.vxPhi} can be identified as follows. As one approaches the phase boundary for band 2 activity, $\xi_2^2$ increases linearly and so does the size of a vortex in that band. This is illustrated in Fig.~\ref{fig.vxPhi} (b) and (c), where we plot the Cooper pair distributions of band 1 in the (1,0) state and band 2 in the (0,1) state at a fixed applied field and different values of $(v_1/v_2)^2$. As an estimate, we define the vortex size as the contour where the absolute value of the order parameter recovers by 50\% its maximum value in the zero coupling limit, which is given by $\psi_{j0} = W\tau_j^{1/2}$. Within this criterion, one can clearly observe the expansion of the band-2 vortex in the (0,1) state as one approaches the phase boundary. Such an increase in the characteristic length of density variations in band 2 induces a decrease in the energy barriers in this band. Hence, when approaching this phase boundary, all processes involving entry or exit of a vortex in band 2 should take place in advance (at a lower field for increasing $\Phi$ and higher field for decreasing $\Phi$). On the other hand, since $\xi_1$ is fixed, one would expect no change in the size of a vortex in band 1 [which is indeed approximately the case as observed in Fig.~\ref{fig.vxPhi} (c)] and thereby no important change in the energy barriers of that band. In this case, processes involving entry or exit of vortices in band 1 are expected to take place at field values that do not change considerably with the parameter $(v_1/v_2)^2$. 

The scenario explained above  has an immediate consequence on how the supercooling, $\Phi_{sc}$, and superheating, $\Phi_{sh}$, fields of each fractional state changes with $(v_1/v_2)^2$. In general, by sweeping the field up and down, both $(L,L+1)$ and $(L+1,L)$ states will decay to $(L+1,L+1)$ at the corresponding superheating field in the upward sweep or to $(L,L)$ at the corresponding supercooling field in the downward sweep. When the original state is $(L,L+1)$,  the $(L+1,L+1)$ state is achieved with the entry of a vortex in band 1 and $\Phi_{sh}$ is essentially constant, while $(L,L)$ is obtained via the escape of a vortex from band 2, thus pushing $\Phi_{sc}$ to higher values as approaching band-2 phase boundary. Conversely, when $(L+1,L)$ is the original state, $\Phi_{sh}$ is pushed to lower values, because of the advance entry of a vortex in band 2, while $\Phi_{sc}$ keeps essentially constant, once only the energy barrier of band 1 is probed.

\subsection{Effect of the interband coupling}\label{ssec.coupling}

\begin{figure}[t]
\centering
  \includegraphics[width=0.95\columnwidth]{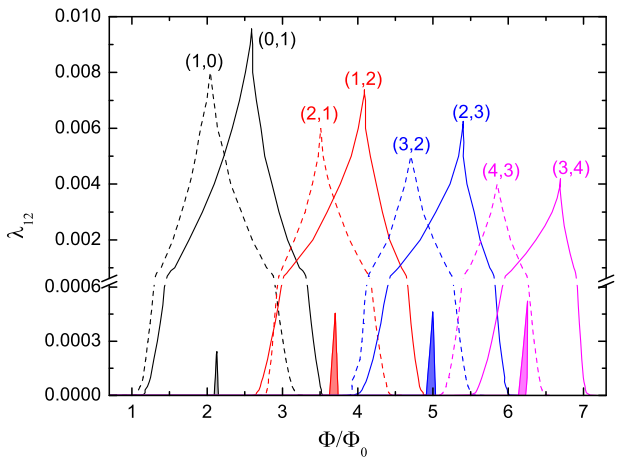}
  \caption{(Color online.) The stability regions of fractional $(L_1,L_2\neq L_1)$ states in the plane defined by the inter band coupling coefficient $\lambda_{12}$ and the magnetic flux $\Phi$ through the sample area. In shaded regions fractional states
represent the thermodynamic equilibrium of the studied system.}
  \label{fig.l12xPhi} 
\end{figure}

The effect of Josephson coupling on the stability of fractional states has been studied in Ref.~\onlinecite{Chibotaru2010} and, in more detail, in Ref.~\onlinecite{Geurts2010}. In these works the Josephson coupling strength $\Gamma$ was treated phenomenologically, without taking into account that the microscopic, interband coupling constant $\lambda_{12}$ can also exert an important influence on  $T_{cj}$ and $T_c$, specially for higher coupling values. In this subection, we investigate the stability regions of fractional states in the $\lambda_{12}$-$\Phi$ plane in consonance with the microscopic background and extend the analysis to the cases of $(L+1,L)$ configurations. Our main results are featured in Fig~\ref{fig.l12xPhi}.
The main trends shown in this figure are similar to those presented in Ref.~\onlinecite{Geurts2010}: shrinking of the stability regions as $\lambda_{12}$ increases and fractional states of lower vorticity are capable of surviving to stronger coupling. Inverted fractional states also present similar behavior. However, their regions of stability are displaced to lower field values. This can be understood in light of our explanation of the supercooling and superheating processes of $(L+1,L)$ and $(L,L+1)$ states (c.f. Sec.~\ref{ssec.coupling}). Here, band 2 is always the weaker band. Therefore, $\Phi_{sh}(L+1,L)$ should be \emph{advanced} to a lower value because it involves vortex entry in band 2, while $\Phi_{sc}(L+1,L)$ should be \emph{delayed} to a lower value since it involves escape of a vortex from the stronger band 1.

Another particularity of $(L+1,L)$ states is that, because of their higher energy, they are less resilient than $(L,L+1)$ states. In fact, for the set of parameters considered here, $(L+1,L)$ states can never be found in thermodynamic equilibrium. In contrast, regions of thermodynamic equilibrium for $(L,L+1)$ can be observed at very low values of $\lambda_{12}$, as can be observed in Fig.~\ref{fig.l12xPhi}, where these regions are represented by shaded areas.

\section{Influence of an off-plane magnetic dot}
\label{sec.MD}

In this section, we analyze the effects of an \emph{inhomogeneous} field on the vortex structures observed in the previous section. We start with particular microscopic parameters for the superconducting disk ($\lambda_{11}=2.415$, $\lambda_{22}=1.211$, $\lambda_{12}=0.001$, $v_1/v_2=1.225$ and $n_1=1-n_2=0.343$) such that fractional states are accessible. The inhomogeneous field is provided by a cylindrical magnetic dot, placed coaxially with the superconducting disk and having a permanent, homogeneous off-plane magnetization given by $\vec{M}=\hat{z}\Phi_M/\pi R^2$. (Here, $\Phi_M$ is not to be confounded with the flux generated by the magnetic dot through the superconductor. Rather, it is only a convenient way to express magnetization in the same units as magnetic field.) To simplify our analysis we chose geometric parameters such that the stray fields generated by the magnetic dot at the superconducting disk have a flat distribution. This is accomplished by a magnetic cylinder of radius $R_{MD}=0.5R = 5$, height $h=4.8$, and placed a distance $l=0.7$ above the superconducting disk (with assumed oxide layer in between, to avoid proximity effect). 

\begin{figure}[b]
\centering
  \includegraphics[width=0.95\columnwidth]{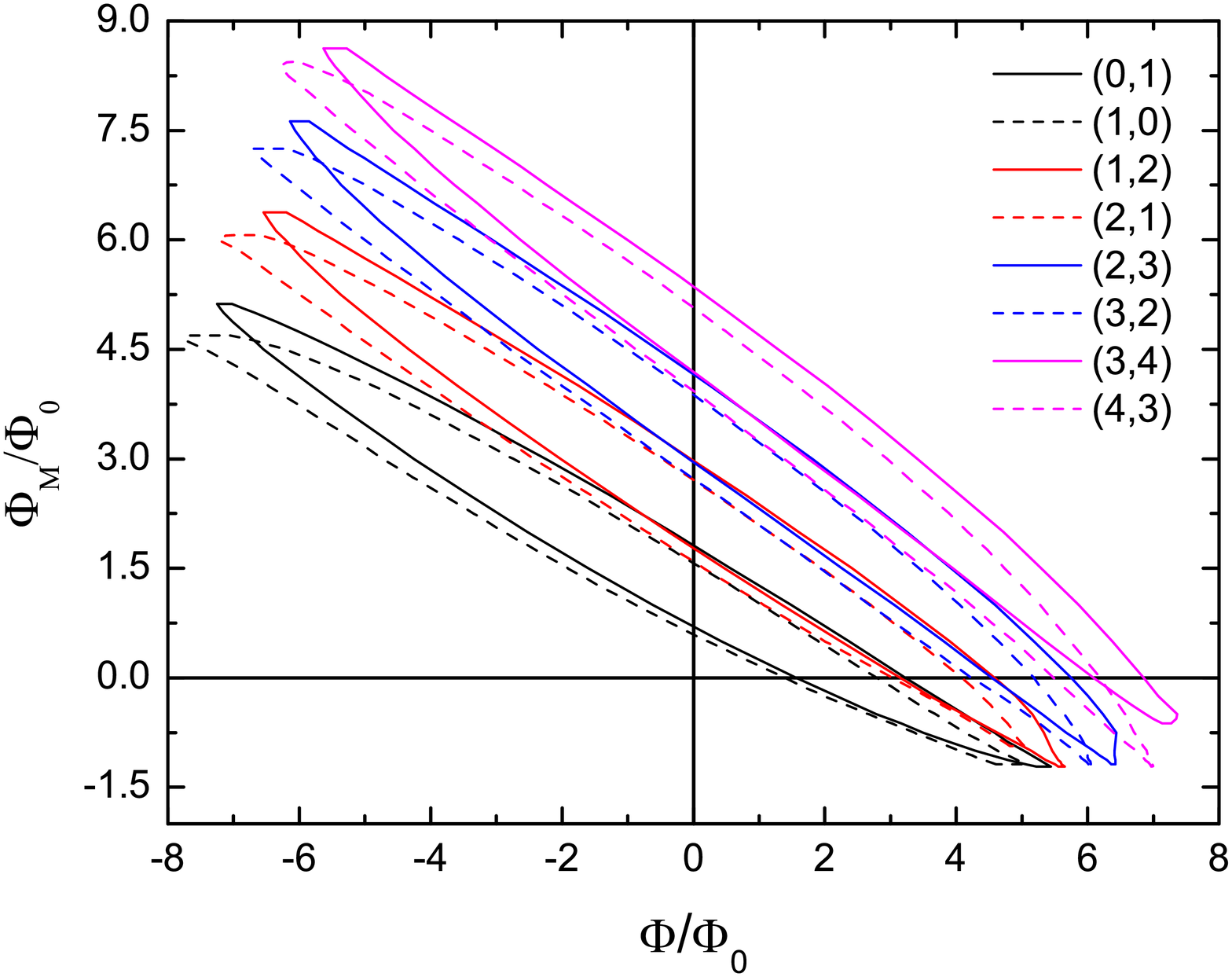}
  \caption{(Color online.) $\Phi_M$-$\Phi$ diagram showing the stability regions of fractional $(L_1,L_2\neq L_1)$ states in the plane defined by the magnetization of the magnetic dot and the external magnetic field, both multiplied by the superconductor area. }
  \label{fig.MxPhi} 
\end{figure}
\begin{figure*}[bt]
\centering
  \includegraphics[width=1.5\columnwidth]{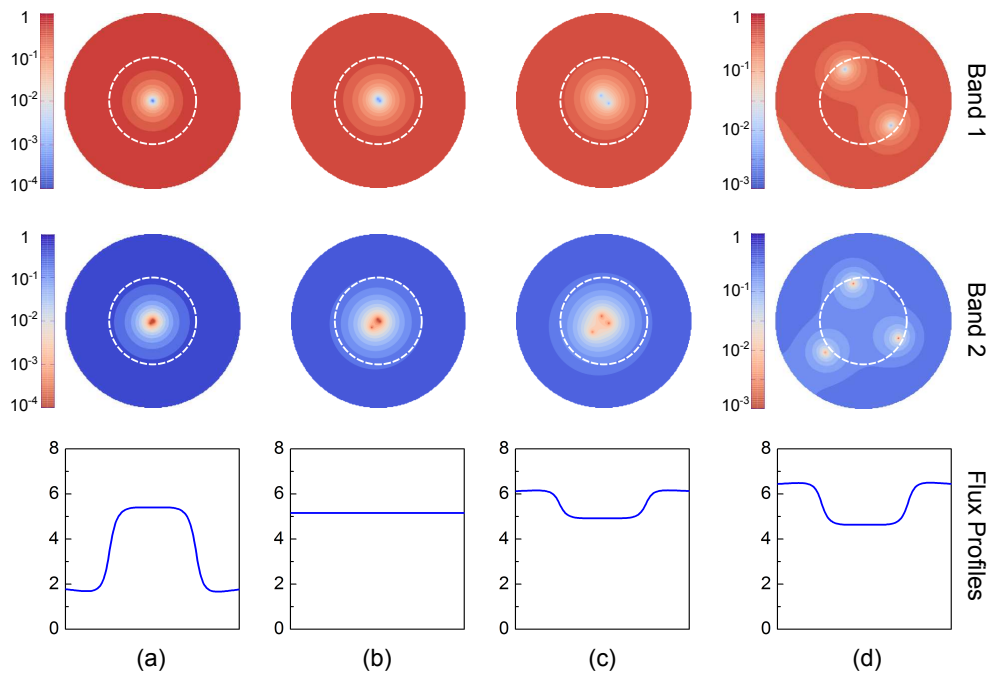}
  \caption{(Color online.) Absolute value of the order parameters and external field profiles (provided by the magnetic dot and homogeneous field source) for the (2,3) state at the median of its stability region for (a) $\Phi_M = 2.25$ and $\Phi = 2.0$, (b)  $\Phi_M = 0$ and $\Phi = 5.15$, (c) $\Phi_M = -0.75$ and $\Phi = 6.05$, and (d) $\Phi_M = -1.125$ and $\Phi = 6.33$. The order parameter of each band is normalized by its respective zero-coupling maximum value and plotted in log scale for better visualization. Dashed circles represent the edge of the magnetic dot. }
  \label{fig.ConfM} 
\end{figure*}

Fig.~\ref{fig.MxPhi} illustrates the evolution of the stability regions of $(L,L+1)$ and $(L+1,L)$ states (for $0\leq L<4$) when the magnetization of the disk is changed. It is promptly observed that the stability of all fractional states can be \emph{considerably enhanced} by the presence of the magnetic dot. This happens for moderate positive values of $\Phi_M$. Too large positive values tend to drive the entire system to the normal state, which of course also destroys the fractional states. On the other hand, negative $\Phi_M$ is deleterious to fractional states because in that case the central region of the superconducting disk encloses less flux, which disables it to host vortices in that area -- while disk periphery is in any case not favorable for vortices to reside.  Therefore, as the external field in this region is fully compensated, the Meissner state becomes predominant. Upon further decreasing $\Phi_M$ to even more negative values, the system becomes susceptible to the appearance of integer and fractional \emph{antivortex} states. The first fractional states appearing are $(0,-1)$ and $(-1,0)$, followed by $(-1,-2)$ and $(-2,-1)$, and so on. The stability regions of such fractional antivortex states correspond to a rotation of the stability regions of their counterparts with positive vorticity, shown in Fig.~\ref{fig.MxPhi}, by 180$^\circ$ with respect to the origin. For the parameters used in our simulations, we did not detect any states where vortices and antivortices coexist in neither the same band nor in different bands, though they are expected to appear for larger samples as already demonstrated in single band superconductors.\cite{Milo2003,Milo2007,Bending2007}

The role of magnetic moment orientation can be better understood by analyzing the field profiles for different $\Phi_M$ and $\Phi$ and the corresponding vortex configurations. For such an analysis, we focus on a particular fractional state, (2,3), and study the evolution of the Cooper-pair density when $\Phi_M$ and $\Phi$ are simultaneously changed in a way as to keep the (2,3) state at the median of its stability field range for each magnetization value. The results are presented in Fig.~\ref{fig.ConfM} where field profiles and Cooper-pair distributions of both bands are shown for the cases $\Phi_M/\Phi_0 = 2.25$, 0, -0.75, and -1.125. In all cases the local flux in the central region of the superconducting disk, where the vortices sit, has approximately the same value, $\sim 5\Phi_0$, which is roughly the flux necessary to stabilize state (2,3) for $\Phi_M=0$.  Therefore, the energy barrier for vortex escape is essentially the same in all cases in such a way that the supercooling field is expected to be an approximately linearly decreasing function of $\Phi_M$. On the other hand, for a new vortex to come in, it has to probe the energy barrier near the superconductor edge, where the flux profile is very sensible to the magnetization of the dot. For positive $\Phi_M$, screening currents are weaker near the sample edges making it more difficult for a new vortex to enter. Therefore the energy barrier for vortex entrance increases (with respect to the median of the stability range) with $\Phi_M$. Hence, superheating is prolonged to higher field values, thereby providing the broadening of the stability range of a given vortex configuration. Note that this picture holds for any vortex arrangements, including integer states. 

A striking feature that can be observed in Fig.~\ref{fig.ConfM} is the expansion of vortex configurations when $\Phi_M$ is decreased down to negative values. Expansion of very compact vortex configurations with the help of ferromagnetic dots has been anticipated to occur for single-band mesoscopic squares in Ref. \onlinecite{Carballeira2005} In the present case, , we demonstrate that this effect can also be applied to multiband materials, thus providing an invaluable tool for experimental visualization of fractional vortex states. This expansion occurs because the decrease of the magnetic flux in the central region of the disk reduces the confining power of the screening currents and allows vortices to repel each other further from the center of the sample. Further reduction of $\Phi_M$ will lead to the sequential expulsion of vortices until the Meissner state is reached. 

\begin{figure}[bt]
\centering
  \includegraphics[width=\columnwidth]{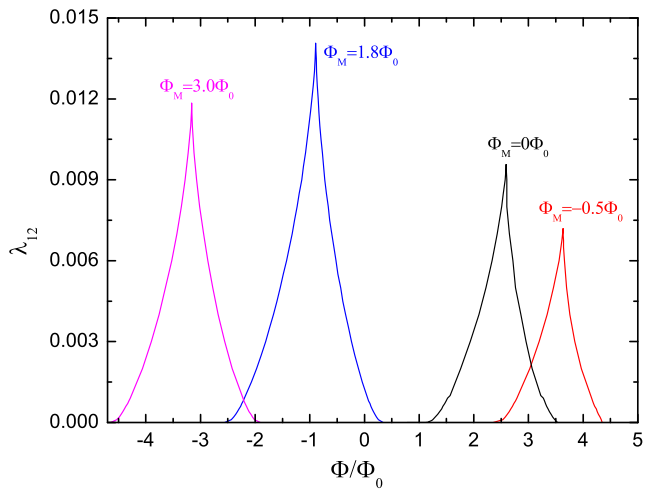}
  \caption{(Color online.) Effect of the magnetic dot magnetization on the stability of state (0,1) in the $\lambda_{12}$-$\Phi$ plane for different values of the reduced magnetization $\Phi_M$. }
  \label{fig.l12MD} 
\end{figure}

In addition to discussing the influence of the magnetic dot on the stability regions for given interband coupling, it is also important to analyze what happens when $\lambda_{12}$ is varied. Fig.~\ref{fig.l12MD}  
demonstrates that not only the range of stability of fractional states can be considerably enhanced but the maximum value of $\lambda_{12}$ in which a given fractional state is stable increases as well. That is to say, \emph{a properly magnetized magnetic dot placed close to a mesoscopic two-band superconductor can optimally stabilize fractional vortex states}, allowing them to be observed even in stronger intraband-coupling materials where they would be otherwise forbidden in a homogeneous field. For the $(0,1)$ case illustrated in Fig.~\ref{fig.l12MD} we found that the optimum magnetization is $\Phi_M \sim 1.8$, for which the maximum value of $\lambda_{12}$ where state (0,1) can be stabilized is enhanced by $\sim$ 45\% compared to $\Phi_M=0$ case. We believe that these values can be considerably improved by exploring other configuration of the magnetic dot. However, the search of the optimum configuration is beyond the scope of this work; we restrict ourselves here to the proof-of-concept.  Although we present here only the $(0,1)$ case, we verified that the same conclusions apply to other fractional states, but of course with different optimum magnetization values.

\section{conclusions}

In summary, we have analyzed the stability of fractional vortex states in a mesoscopic superconductor with respect to changing its microscopic material parameters and to the magnetization state of a magnetic dot placed nearby. By sweeping a broad range of values of the partial density of states and of the Fermi velocities for a given coupling matrix, we have determined the stability regions of many different fractional vortex states. We could identify two main groups of such solutions: the first one comprises those states where the dominating band, i.e. that one with the highest $T_{cj}$, has a lower total winding number; the second group comprises states where the dominating band has the highest winding number. As expected, states of the first group have, in general, lower energy and are easier to find because the dominating band has a natural tendency to be more refractory to vortices. In addition, they can be found in thermodynamic equilibrium for the case of weak Josephson coupling, as previously reported in the literature.\cite{Chibotaru2010,Geurts2010} However, the states in each of these groups can exchange roles smoothly by sweeping the partial density of states. We have also demonstrated that the healing length of a band at zero temperature, defined as $\xi_j=\hbar v_j/\sqrt{6W}$, is closely connected to the size of a vortex in that band and thereby to the ability of the surface barrier to superheat or supercool fractional states in applied magnetic field. 

We have further investigated the properties of fractional vortex states in \emph{inhomogeneous} magnetic field. In particular, we exposed the superconducting disk to the stray field of a magnetic dot, placed coaxially above the sample. We demonstrated that such a configuration is capable of augmenting the stability region of fractional states, at least when the dot magnetization is parallel to the external homogeneous field. Moreover, the maximum value of interband coupling for which a given fractional state is stable can be increased considerably and reaches an optimal value at a well defined magnetization of the dot. This is an important result because fractional states are usually possible only in materials with very weak interband coupling. In other words, the here proposed idea enables observation of fractional vortex states in a considerably wider range of multiband materials.

Finally, we point out that, to date, experiments determining vortex structures in two-band superconductors have been restricted to bulk single crystals. However, mesoscopic systems present a much richer variety of vortex states, including non-composite fractional vortices, which represent a clear, direct evidence of competing length scales in multiband materials. We believe our findings contribute to the construction of a roadmap guiding further search of suitable materials for imaging these exotic vortex structures. In this respect, further investigation on other geometries of the mesoscopic superconductor and the magnetic dot aiming at the stabilization of fractional vortices in wider parameter ranges would be very much welcome.

\acknowledgements

We thank Eric B. Claude,  Miguel A. Zorro, and Rog\'erio M. da Silva for assistance in the development of the numerical code used in our simulations. This work was supported by the Brazilian science agencies CNPq and FACEPE, by the FACEPE/CNPq-PRONEX program, under Grant No. APQ-0589-1.05/08, and by CNPq-FWO Brazil-Flanders cooperation program. MVM acknowledges support from the CAPES-PVE program.

\bibliographystyle{apsrev}

\begin{thebibliography}{31}
\expandafter\ifx\csname natexlab\endcsname\relax\def\natexlab#1{#1}\fi
\expandafter\ifx\csname bibnamefont\endcsname\relax
  \def\bibnamefont#1{#1}\fi
\expandafter\ifx\csname bibfnamefont\endcsname\relax
  \def\bibfnamefont#1{#1}\fi
\expandafter\ifx\csname citenamefont\endcsname\relax
  \def\citenamefont#1{#1}\fi
\expandafter\ifx\csname url\endcsname\relax
  \def\url#1{\texttt{#1}}\fi
\expandafter\ifx\csname urlprefix\endcsname\relax\def\urlprefix{URL }\fi
\providecommand{\bibinfo}[2]{#2}
\providecommand{\eprint}[2][]{\url{#2}}

\bibitem[{\citenamefont{Leggett}(2008)}]{Leggett2008}
\bibinfo{author}{\bibfnamefont{A.~J.} \bibnamefont{Leggett}},
  \emph{\bibinfo{title}{Quantum Liquids}} (\bibinfo{publisher}{Oxford
  University Press, New York}, \bibinfo{year}{2008}).

\bibitem[{\citenamefont{Abrikosov}(1957)}]{Abrikosov57}
\bibinfo{author}{\bibfnamefont{A.}~\bibnamefont{Abrikosov}},
  \bibinfo{journal}{Sov. Phys. JETP} \textbf{\bibinfo{volume}{5}},
  \bibinfo{pages}{1174} (\bibinfo{year}{1957}).

\bibitem[{\citenamefont{Babaev}(2002)}]{Babaev2002}
\bibinfo{author}{\bibfnamefont{E.}~\bibnamefont{Babaev}},
  \bibinfo{journal}{Phys. Rev. Lett.} \textbf{\bibinfo{volume}{89}},
  \bibinfo{pages}{067001} (\bibinfo{year}{2002}).

\bibitem[{\citenamefont{Komendov\'{a} et~al.}(2011)\citenamefont{Komendov\'{a},
  Milo\v{s}evi\'{c}, Shanenko, and Peeters}}]{Komendova2011}
\bibinfo{author}{\bibfnamefont{L.}~\bibnamefont{Komendov\'{a}}},
  \bibinfo{author}{\bibfnamefont{M.~V.} \bibnamefont{Milo\v{s}evi\'{c}}},
  \bibinfo{author}{\bibfnamefont{A.~A.} \bibnamefont{Shanenko}},
  \bibnamefont{and} \bibinfo{author}{\bibfnamefont{F.~M.}
  \bibnamefont{Peeters}}, \bibinfo{journal}{Phys. Rev. B}
  \textbf{\bibinfo{volume}{84}}, \bibinfo{pages}{064522}
  (\bibinfo{year}{2011}).

\bibitem[{\citenamefont{Silaev and Babaev}(2012)}]{Silaev2012}
\bibinfo{author}{\bibfnamefont{M.}~\bibnamefont{Silaev}} \bibnamefont{and}
  \bibinfo{author}{\bibfnamefont{E.}~\bibnamefont{Babaev}},
  \bibinfo{journal}{Phys. Rev. B} \textbf{\bibinfo{volume}{85}},
  \bibinfo{pages}{134514} (\bibinfo{year}{2012}).

\bibitem[{\citenamefont{Silaev and Babaev}(2011)}]{Silaev2011}
\bibinfo{author}{\bibfnamefont{M.}~\bibnamefont{Silaev}} \bibnamefont{and}
  \bibinfo{author}{\bibfnamefont{E.}~\bibnamefont{Babaev}},
  \bibinfo{journal}{Phys. Rev. B} \textbf{\bibinfo{volume}{84}},
  \bibinfo{pages}{094515} (\bibinfo{year}{2011}).

\bibitem[{\citenamefont{Babaev and Speight}(2005)}]{Babaev2005}
\bibinfo{author}{\bibfnamefont{E.}~\bibnamefont{Babaev}} \bibnamefont{and}
  \bibinfo{author}{\bibfnamefont{M.}~\bibnamefont{Speight}},
  \bibinfo{journal}{Phys. Rev. B} \textbf{\bibinfo{volume}{72}},
  \bibinfo{pages}{180502} (\bibinfo{year}{2005}).

\bibitem[{\citenamefont{Moshchalkov et~al.}(2009)\citenamefont{Moshchalkov,
  Menghini, Nishio, Chen, Silhanek, Dao, Chibotaru, Zhigadlo, and
  Karpinski}}]{Moshchalkov2009}
\bibinfo{author}{\bibfnamefont{V.}~\bibnamefont{Moshchalkov}},
  \bibinfo{author}{\bibfnamefont{M.}~\bibnamefont{Menghini}},
  \bibinfo{author}{\bibfnamefont{T.}~\bibnamefont{Nishio}},
  \bibinfo{author}{\bibfnamefont{Q.~H.} \bibnamefont{Chen}},
  \bibinfo{author}{\bibfnamefont{A.~V.} \bibnamefont{Silhanek}},
  \bibinfo{author}{\bibfnamefont{V.~H.} \bibnamefont{Dao}},
  \bibinfo{author}{\bibfnamefont{L.~F.} \bibnamefont{Chibotaru}},
  \bibinfo{author}{\bibfnamefont{N.~D.} \bibnamefont{Zhigadlo}},
  \bibnamefont{and}
  \bibinfo{author}{\bibfnamefont{J.}~\bibnamefont{Karpinski}},
  \bibinfo{journal}{Phys. Rev. Lett.} \textbf{\bibinfo{volume}{102}},
  \bibinfo{pages}{117001} (\bibinfo{year}{2009}).

\bibitem[{\citenamefont{Dao et~al.}(2011)\citenamefont{Dao, Chibotaru, Nishio,
  and Moshchalkov}}]{Dao2011}
\bibinfo{author}{\bibfnamefont{V.~H.} \bibnamefont{Dao}},
  \bibinfo{author}{\bibfnamefont{L.~F.} \bibnamefont{Chibotaru}},
  \bibinfo{author}{\bibfnamefont{T.}~\bibnamefont{Nishio}}, \bibnamefont{and}
  \bibinfo{author}{\bibfnamefont{V.~V.} \bibnamefont{Moshchalkov}},
  \bibinfo{journal}{Phys. Rev. B} \textbf{\bibinfo{volume}{83}},
  \bibinfo{pages}{020503} (\bibinfo{year}{2011}).

\bibitem[{\citenamefont{Babaev et~al.}(2009)\citenamefont{Babaev,
  J\"{a}ykk\"{a}, and Speight}}]{Babaev2009}
\bibinfo{author}{\bibfnamefont{E.}~\bibnamefont{Babaev}},
  \bibinfo{author}{\bibfnamefont{J.}~\bibnamefont{J\"{a}ykk\"{a}}},
  \bibnamefont{and} \bibinfo{author}{\bibfnamefont{M.}~\bibnamefont{Speight}},
  \bibinfo{journal}{Phys. Rev. Lett.} \textbf{\bibinfo{volume}{103}},
  \bibinfo{pages}{237002} (\bibinfo{year}{2009}).

\bibitem[{\citenamefont{Cho and Zhang}(2008)}]{ChoZhang2008}
\bibinfo{author}{\bibfnamefont{Y.}~\bibnamefont{Cho}} \bibnamefont{and}
  \bibinfo{author}{\bibfnamefont{P.}~\bibnamefont{Zhang}},
  \bibinfo{journal}{Eur. Phys. J. B} \textbf{\bibinfo{volume}{65}},
  \bibinfo{pages}{155} (\bibinfo{year}{2008}).

\bibitem[{\citenamefont{Chibotaru et~al.}(2007)\citenamefont{Chibotaru, Dao,
  and Ceulemans}}]{Chibotaru2007}
\bibinfo{author}{\bibfnamefont{L.~F.} \bibnamefont{Chibotaru}},
  \bibinfo{author}{\bibfnamefont{V.~H.} \bibnamefont{Dao}}, \bibnamefont{and}
  \bibinfo{author}{\bibfnamefont{A.}~\bibnamefont{Ceulemans}},
  \bibinfo{journal}{EPL} \textbf{\bibinfo{volume}{78}}, \bibinfo{pages}{47001}
  (\bibinfo{year}{2007}).

\bibitem[{\citenamefont{Chibotaru and Dao}(2010)}]{Chibotaru2010}
\bibinfo{author}{\bibfnamefont{L.~F.} \bibnamefont{Chibotaru}}
  \bibnamefont{and} \bibinfo{author}{\bibfnamefont{V.~H.} \bibnamefont{Dao}},
  \bibinfo{journal}{Phys. Rev. B} \textbf{\bibinfo{volume}{81}},
  \bibinfo{pages}{020502(R)} (\bibinfo{year}{2010}).

\bibitem[{\citenamefont{Geurts et~al.}(2010)\citenamefont{Geurts,
  Milo\v{s}evi\'{c}, and Peeters}}]{Geurts2010}
\bibinfo{author}{\bibfnamefont{R.}~\bibnamefont{Geurts}},
  \bibinfo{author}{\bibfnamefont{M.~V.} \bibnamefont{Milo\v{s}evi\'{c}}},
  \bibnamefont{and} \bibinfo{author}{\bibfnamefont{F.~M.}
  \bibnamefont{Peeters}}, \bibinfo{journal}{Phys. Rev. B}
  \textbf{\bibinfo{volume}{81}}, \bibinfo{pages}{214514}
  (\bibinfo{year}{2010}).

\bibitem[{\citenamefont{Chaves et~al.}(2011)\citenamefont{Chaves, Komendov\'a,
  Milo\v{s}evi\'{c}, Andrade, Farias, and Peeters}}]{Chaves2011}
\bibinfo{author}{\bibfnamefont{A.}~\bibnamefont{Chaves}},
  \bibinfo{author}{\bibfnamefont{L.}~\bibnamefont{Komendov\'a}},
  \bibinfo{author}{\bibfnamefont{M.~V.} \bibnamefont{Milo\v{s}evi\'{c}}},
  \bibinfo{author}{\bibfnamefont{J.~S.} \bibnamefont{Andrade}},
  \bibinfo{author}{\bibfnamefont{G.~A.} \bibnamefont{Farias}},
  \bibnamefont{and} \bibinfo{author}{\bibfnamefont{F.~M.}
  \bibnamefont{Peeters}}, \bibinfo{journal}{Phys. Rev. B}
  \textbf{\bibinfo{volume}{83}}, \bibinfo{pages}{214523}
  (\bibinfo{year}{2011}).

\bibitem[{\citenamefont{Zhitomirsky and Dao}(2004)}]{Zhitomirsky2004}
\bibinfo{author}{\bibfnamefont{M.~E.} \bibnamefont{Zhitomirsky}}
  \bibnamefont{and} \bibinfo{author}{\bibfnamefont{V.-H.} \bibnamefont{Dao}},
  \bibinfo{journal}{Phys. Rev. B} \textbf{\bibinfo{volume}{69}},
  \bibinfo{pages}{054508} (\bibinfo{year}{2004}).

\bibitem[{\citenamefont{Kogan and Schmalian}(2011)}]{Kogan2011}
\bibinfo{author}{\bibfnamefont{V.~G.} \bibnamefont{Kogan}} \bibnamefont{and}
  \bibinfo{author}{\bibfnamefont{J.}~\bibnamefont{Schmalian}},
  \bibinfo{journal}{Phys. Rev. B} \textbf{\bibinfo{volume}{83}},
  \bibinfo{pages}{054515} (\bibinfo{year}{2011}).

\bibitem[{\citenamefont{Komendov\'{a} et~al.}(2012)\citenamefont{Komendov\'{a},
  Chen, Shanenko, Milo\v{s}evi\'{c}, and Peeters}}]{Komendova2012}
\bibinfo{author}{\bibfnamefont{L.}~\bibnamefont{Komendov\'{a}}},
  \bibinfo{author}{\bibfnamefont{Y.}~\bibnamefont{Chen}},
  \bibinfo{author}{\bibfnamefont{A.~A.} \bibnamefont{Shanenko}},
  \bibinfo{author}{\bibfnamefont{M.~V.} \bibnamefont{Milo\v{s}evi\'{c}}},
  \bibnamefont{and} \bibinfo{author}{\bibfnamefont{F.~M.}
  \bibnamefont{Peeters}}, \bibinfo{journal}{Phys. Rev. Lett.}
  \textbf{\bibinfo{volume}{108}}, \bibinfo{pages}{207002}
  (\bibinfo{year}{2012}).

\bibitem[{\citenamefont{Shanenko et~al.}(2011)\citenamefont{Shanenko,
  Milo\v{s}evi\'{c}, Peeters, and Vagov}}]{Shanenko2011}
\bibinfo{author}{\bibfnamefont{A.~A.} \bibnamefont{Shanenko}},
  \bibinfo{author}{\bibfnamefont{M.~V.} \bibnamefont{Milo\v{s}evi\'{c}}},
  \bibinfo{author}{\bibfnamefont{F.~M.} \bibnamefont{Peeters}},
  \bibnamefont{and} \bibinfo{author}{\bibfnamefont{A.~V.} \bibnamefont{Vagov}},
  \bibinfo{journal}{Phys. Rev. Lett.} \textbf{\bibinfo{volume}{106}},
  \bibinfo{pages}{047005} (\bibinfo{year}{2011}).

\bibitem[{\citenamefont{Vagov et~al.}(2012)\citenamefont{Vagov, Shanenko,
  Milo\v{s}evi\'{c}, Axt, and Peeters}}]{Vagov2012}
\bibinfo{author}{\bibfnamefont{A.~V.} \bibnamefont{Vagov}},
  \bibinfo{author}{\bibfnamefont{A.~A.} \bibnamefont{Shanenko}},
  \bibinfo{author}{\bibfnamefont{M.~V.} \bibnamefont{Milo\v{s}evi\'{c}}},
  \bibinfo{author}{\bibfnamefont{V.~M.} \bibnamefont{Axt}}, \bibnamefont{and}
  \bibinfo{author}{\bibfnamefont{F.~M.} \bibnamefont{Peeters}},
  \bibinfo{journal}{Phys. Rev. B} \textbf{\bibinfo{volume}{85}},
  \bibinfo{pages}{014502} (\bibinfo{year}{2012}).

\bibitem[{\citenamefont{Winiecki and Adams}(2002)}]{Winiecki2002}
\bibinfo{author}{\bibfnamefont{T.}~\bibnamefont{Winiecki}} \bibnamefont{and}
  \bibinfo{author}{\bibfnamefont{C.}~\bibnamefont{Adams}},
  \bibinfo{journal}{Journal of Computational Physics}
  \textbf{\bibinfo{volume}{179}}, \bibinfo{pages}{127} (\bibinfo{year}{2002}).

\bibitem[{\citenamefont{Brandt and Das}(2011)}]{Brandt2011}
\bibinfo{author}{\bibfnamefont{E.~H.} \bibnamefont{Brandt}} \bibnamefont{and}
  \bibinfo{author}{\bibfnamefont{M.~P.} \bibnamefont{Das}},
  \bibinfo{journal}{J. Supercond. Nov. Magn.} \textbf{\bibinfo{volume}{24}},
  \bibinfo{pages}{57} (\bibinfo{year}{2011}).

\bibitem[{\citenamefont{Geim et~al.}(1997)\citenamefont{Geim, Grigorieva,
  Dubonos, Lok, Maan, Filippov, and Peeters}}]{Geim97}
\bibinfo{author}{\bibfnamefont{A.~K.} \bibnamefont{Geim}},
  \bibinfo{author}{\bibfnamefont{I.~V.} \bibnamefont{Grigorieva}},
  \bibinfo{author}{\bibfnamefont{S.~V.} \bibnamefont{Dubonos}},
  \bibinfo{author}{\bibfnamefont{J.~G.~S.} \bibnamefont{Lok}},
  \bibinfo{author}{\bibfnamefont{J.~C.} \bibnamefont{Maan}},
  \bibinfo{author}{\bibfnamefont{A.~E.} \bibnamefont{Filippov}},
  \bibnamefont{and} \bibinfo{author}{\bibfnamefont{F.~M.}
  \bibnamefont{Peeters}}, \bibinfo{journal}{Nature}
  \textbf{\bibinfo{volume}{390}}, \bibinfo{pages}{259} (\bibinfo{year}{1997}).

\bibitem[{\citenamefont{Geim et~al.}(1998)\citenamefont{Geim, Dubonos, Lok,
  Henini, and Maan}}]{Geim98}
\bibinfo{author}{\bibfnamefont{A.~K.} \bibnamefont{Geim}},
  \bibinfo{author}{\bibfnamefont{S.~V.} \bibnamefont{Dubonos}},
  \bibinfo{author}{\bibfnamefont{J.~G.~S.} \bibnamefont{Lok}},
  \bibinfo{author}{\bibfnamefont{M.}~\bibnamefont{Henini}}, \bibnamefont{and}
  \bibinfo{author}{\bibfnamefont{J.~C.} \bibnamefont{Maan}},
  \bibinfo{journal}{Nature} \textbf{\bibinfo{volume}{396}},
  \bibinfo{pages}{144} (\bibinfo{year}{1998}).

\bibitem[{\citenamefont{Schweigert and Peeters}(1999)}]{Schw1999}
\bibinfo{author}{\bibfnamefont{V.~A.} \bibnamefont{Schweigert}}
  \bibnamefont{and} \bibinfo{author}{\bibfnamefont{F.~M.}
  \bibnamefont{Peeters}}, \bibinfo{journal}{Phys. Rev. Lett.}
  \textbf{\bibinfo{volume}{83}}, \bibinfo{pages}{2409} (\bibinfo{year}{1999}).

\bibitem[{\citenamefont{Palacios}(2000)}]{Palacios2000}
\bibinfo{author}{\bibfnamefont{J.~J.} \bibnamefont{Palacios}},
  \bibinfo{journal}{Phys. Rev. Lett.} \textbf{\bibinfo{volume}{84}},
  \bibinfo{pages}{1796} (\bibinfo{year}{2000}).

\bibitem[{\citenamefont{Baelus et~al.}(2001)\citenamefont{Baelus, Peeters, and
  Schweigert}}]{Baelus2001}
\bibinfo{author}{\bibfnamefont{B.~J.} \bibnamefont{Baelus}},
  \bibinfo{author}{\bibfnamefont{F.~M.} \bibnamefont{Peeters}},
  \bibnamefont{and} \bibinfo{author}{\bibfnamefont{V.~A.}
  \bibnamefont{Schweigert}}, \bibinfo{journal}{Phys. Rev. B}
  \textbf{\bibinfo{volume}{63}}, \bibinfo{pages}{144517}
  (\bibinfo{year}{2001}).

\bibitem[{\citenamefont{Milo\v{s}evi\'{c} and Peeters}(2003)}]{Milo2003}
\bibinfo{author}{\bibfnamefont{M.~V.} \bibnamefont{Milo\v{s}evi\'{c}}}
  \bibnamefont{and} \bibinfo{author}{\bibfnamefont{F.~M.}
  \bibnamefont{Peeters}}, \bibinfo{journal}{Phys. Rev. B}
  \textbf{\bibinfo{volume}{68}}, \bibinfo{pages}{024509}
  (\bibinfo{year}{2003}).

\bibitem[{\citenamefont{Milo\v{s}evi\'{c}
  et~al.}(2007)\citenamefont{Milo\v{s}evi\'{c}, Berdiyorov, and
  Peeters}}]{Milo2007}
\bibinfo{author}{\bibfnamefont{M.~V.} \bibnamefont{Milo\v{s}evi\'{c}}},
  \bibinfo{author}{\bibfnamefont{G.~R.} \bibnamefont{Berdiyorov}},
  \bibnamefont{and} \bibinfo{author}{\bibfnamefont{F.~M.}
  \bibnamefont{Peeters}}, \bibinfo{journal}{Phys. Rev. B}
  \textbf{\bibinfo{volume}{75}}, \bibinfo{pages}{052502}
  (\bibinfo{year}{2007}).

\bibitem[{\citenamefont{Neal et~al.}(2007)\citenamefont{Neal,
  Milo\v{s}evi\'{c}, Bending, Potenza, Emeterio, and Marrows}}]{Bending2007}
\bibinfo{author}{\bibfnamefont{J.~S.} \bibnamefont{Neal}},
  \bibinfo{author}{\bibfnamefont{M.~V.} \bibnamefont{Milo\v{s}evi\'{c}}},
  \bibinfo{author}{\bibfnamefont{S.~J.} \bibnamefont{Bending}},
  \bibinfo{author}{\bibfnamefont{A.}~\bibnamefont{Potenza}},
  \bibinfo{author}{\bibfnamefont{L.~S.} \bibnamefont{Emeterio}},
  \bibnamefont{and} \bibinfo{author}{\bibfnamefont{C.~H.}
  \bibnamefont{Marrows}}, \bibinfo{journal}{Phys. Rev. Lett.}
  \textbf{\bibinfo{volume}{99}}, \bibinfo{pages}{127001}
  (\bibinfo{year}{2007}).

\bibitem[{\citenamefont{Carballeira et~al.}(2005)\citenamefont{Carballeira,
  Moshchalkov, Chibotaru, and Ceulemans}}]{Carballeira2005}
\bibinfo{author}{\bibfnamefont{C.}~\bibnamefont{Carballeira}},
  \bibinfo{author}{\bibfnamefont{V.~V.} \bibnamefont{Moshchalkov}},
  \bibinfo{author}{\bibfnamefont{L.~F.} \bibnamefont{Chibotaru}},
  \bibnamefont{and}
  \bibinfo{author}{\bibfnamefont{A.}~\bibnamefont{Ceulemans}},
  \bibinfo{journal}{Phys. Rev. Lett.} \textbf{\bibinfo{volume}{95}},
  \bibinfo{pages}{237003} (\bibinfo{year}{2005}).

\end{thebibliography}

\end{document}